\documentclass[twocolumn,aip,floats,showpacs]{revtex4}
\usepackage{graphicx}
\usepackage{amssymb}

\newcommand{\vect}[1]{{\mbox{\boldmath $#1$}}}

\begin{document}

\title{  Spin Dynamics and Multiple Reflections in Ferromagnetic Film in Contact with Normal Metal Layers}


\author{E. \v{S}im\'{a}nek}
 \email{simanek@ucr.edu}

\affiliation{Department of Physics, University of California,
Riverside, CA 92521}

\begin{abstract} Spin dynamics of a metallic ferromagnetic film
imbedded between normal metal layers is studied using the
spin-pumping theory of Tserkovnyak \textit{et al.}[Phys. Rev.
Lett. \textbf{88}, 117601 (2002)]. The scattering matrix for this
structure is obtained using a spin-dependent potential with
quantum well in the ferromagnetic region. Owing to multiple
reflections in the well, the excess Gilbert damping and the
gyromagnetic ratio exhibit quantum oscillations as a function of
the thickness of the ferromagnetic film. The wavelength of the
oscillations is given by the depth of the quantum well. For iron
film imbedded between gold layers, the amplitude of the
oscillations of the Gilbert damping is in an order of magnitude
agreement with the damping observed by Urban \textit{et al.}[Phys.
Rev. Lett. \textbf{87}, 217204 (2001)]. The results are compared
with the linear response theory of Mills [Phys. Rev. B
\textbf{68}, 0144419 (2003)].

\end{abstract}


\maketitle

\section{Introduction}

Spin dynamics in multilayers consisting of a thin ferromagnetic
film (F) imbedded between normal metal films (N) has received much
attention recently.  Ferromagnetic resonance (FMR) experiments on
N/F/N structures show that imbedding produces an additional spin
damping$^{1,2}$.  A possible explanation of these observations is
that the localized spins in the ferromagnet interact, via s-d
exchange, with the conduction electrons in the normal metal$^{3}$.
Due to this interaction, a transfer of angular momentum takes
place from the precessing ferromagnet to the conduction electrons
in the adjacent N layers.  The reactive torque generated by this
transfer acts back on the ferromagnetic magnetization causing an
increased linewidth and a shift of the ferromagnetic resonance
absorption curve.

An interesting formulation of this mechanism has been given in
recent work by Tserkovnyak et al$^4$.  Extending the scattering
approach of parametric charge pumping by Brouwer$^5$ to spin
pumping, these authors calculate the spin current pumped through
the N/F and F/N contacts of the N/F/N structure.  The
modifications of the spin dynamics, due to the induced spin
current, are expressed using the Landau-Lifshitz-Gilbert (LLG)
equation$^{6,7}$

\begin{equation}\label{Eq1}
\frac{d \vect{m}}{d t} = -\gamma\vect{m} \times \vect{H}_{eff} +
\alpha \vect{m} \times \frac{d \vect{m}}{d t}
\end{equation}

where $\vect{H}_{eff}$ is the effective magnetic field, $\vect{m}$
is the unit vector in the direction of the ferromagnetic
magnetization, $\gamma$ is the gyromagnetic ratio, and $\alpha$ is
the Gilbert damping constant.

According to Ref. [4], the spin pumping produces renormalization
of the parameters $\gamma$ and $\alpha$ given by

\begin{equation}\label{Eq2}
\frac{1}{\gamma} = \frac{1}{\gamma_{0}} \big \{1
+\frac{g_{L}\mu_B}{4 \pi M } [A_{i}^{(L)} +A_{i}^{R}]\big \}
\end{equation}

\begin{equation}\label{Eq3}
\alpha =\frac{\gamma}{\gamma_{0}} \big \{\alpha_{0} +
\frac{g_{L}\mu_B}{4 \pi M}[A_{r}^{(L)} + A _{r}^{(R)}] \big\}
\end{equation}

where parameters $\alpha _{0}$ and $\gamma_{0}$ are the bulk
values of $\alpha$ and $\gamma$. $g_{L}$ is the Lande' factor,
$\mu_B$ is the Bohr magneton, and $M$ is the total moment of the
ferromagnetic film.  The superscripts (L) and (R) correspond to
the left (N/F) and right (F/N) interfaces of the N/F/N structure.
The parameters $A_i^{(L,R)}$ and $A_r^{(L,R)}$ stem from the spin
currents pumped into the (L,R) normal layers (leads). These
parameters are expressed in terms of the scattering amplitudes as
follows$^4$

\begin{equation}\label{Eq4}
A_{r} = \frac{1}{2}\sum_{mn}\big \{|r ^{\uparrow}_{mn} - r
^{\downarrow}_{mn}|^{2}+ | t^{'\uparrow}_{mn}-
t^{'\downarrow}_{mn}|^{2} \big \}
\end{equation}

\begin{equation}\label{Eq5}
A_{i} = \textrm{Im}\sum_{mn}\big \{r
^{\uparrow}_{mn}(r^{\downarrow}_{mn}) ^{*} +
t^{'\uparrow}_{mn}(t^{'\downarrow}_{mn}) ^{*} \big \}
\end{equation}

where $(r ^{\uparrow}_{mn}, r^{\downarrow}_{mn})$ and $(t
^{'\uparrow}_{mn}, t ^{'\downarrow}_{mn})$ are the reflection and
transmission amplitudes for electrons with up and down spins. The
expressions (4) and (5) are to be evaluated with transverse modes
$(m,n)$ taken at the Fermi energy.

From Eqs. (2) and (3) we see, in conjunction with Eq. (1), that
the parameter $A_{i}$ is responsible for the shift of the FMR
frequency, whereas $A_{r}$ increases the FMR linewith. For a
symmetric N/F/N structure, we have $A^{(L)}_{i} = A ^{(R)}_{i} = A
_{i} $, and $A^{(L)}_{r} = A ^{(R)}_{r} = A_{r}$.

When the electrons scattered from one interface interfere
incoherently with the other interface, the parameters $A_i^{(L)}$
and $A_r^{(L)}$  contribute independently of the parameters
$A_i^{(R)}$ and $A_r^{(R)}$. As pointed out in Ref. 4, this
applies for ferromagnetic films that are thick compared to the
coherence length $\lambda_{c}\sim (k^{\uparrow}_{F} -
k^{\downarrow}_{F})^{-1}$ where $k^{\uparrow}_{F}$ and
$k^{\downarrow}_{F}$ are the spin-up and spin-down Fermi wave
vectors, respectively. In this case, the parameters $A_i$ and
$A_r$ can be expressed in terms of the mixing conductances of the
F/N and N/F contacts$^8$.

This approximation suppresses the effect of multiple reflections
in ferromagnetic layers that are not much thicker than
$\lambda_{c}$.  The physics is very similar to that of the
Fabry-Perot interferometer in optics.  One manifestation of such
interferences is oscillation of interlayer exchange coupling in
F/N/F multilayers as a function of ferromagnetic-layer thickness.
This effect has been first predicted from numerical calculations
by Barnas$^9$.  Subsequently, Bruno$^{10}$ offered an explanation
based on quantum interference picture.  Results of his work have
been confirmed by Bloemen et al.$^{11}$ who succeeded to see
oscillations of the interlayer exchange coupling in Co/Cu/Co as
the thickness of the Co layers is varied.

The subject of the present work is to apply the spin-pumping
approach to a model which allows multiple reflections in the
ferromagnetic layer to be manifested in the renormalization of the
parameters $\alpha$ and $\gamma$.  Primary motivation for this
investigation comes from the recent work by Mills$^{12}$.

The starting point of Mills work is an approach$^{13}$ which
generalizes the Ruderman-Kittel-Kasuya-Yosida (RKKY) theory to
time-dependent source represented by the precessing vector
$\vect{m}(t)$.  In this approach, the renormalization of the
parameters $\gamma$ and $\alpha$ is found proportional to the
frequency derivative of the real and imaginary parts of the
transverse susceptibility, respectively.

 Mills calculates the transverse susceptibility from a model
 which allows for multiple reflections in the ferromagnetic
layer.  Instead of representing the ferromagnetic layer by two
independently scattering outside interfaces, (as done in Refs. [4]
and [13 ]), the electrons in Ref. [12] are assumed to move in a
spin-dependent potential of the trilayer in the form of a quantum
well. This analysis yields an excess damping parameter
$\alpha'=\alpha-\alpha_0$ which exhibits strong quantum
oscillations with the thickness of the ferromagnetic film.

Ref. [13] and recent work of the present author$^{14}$ show that,
for a ferromagnetic monolayer, the dynamic RKKY approach and the
spin pumping theory yield the same results for the parameter
$\alpha'$. Thus, we anticipate that application of the
spin-pumping method to scattering by a quantum well potential will
produce renormalizations of the parameters $\alpha$ and $\gamma$
which also exhibit oscillations with the thickness of the
ferromagnetic film.

In the present investigation we confirm this anticipation.
However, it turns out that our results are, in several aspects,
different from those of Ref. 12. Primarily, we find that besides
oscillations of $\alpha'$ there are also oscillations of the
$g$-shift with similar period and magnitude. Moreover, the periods
of the oscillations show a dependence on the exchange splitting
that is different from that of Ref. 12.

\section {Scattering by a Square Well}

We consider a trilayer, N/F/N, consisting of a ferromagnetic layer
of thickness $D$ imbedded between normal metal layers, each of
infinite thickness.  The conduction electrons move in a
spin-dependent potential $U_{\alpha}(x)$ where $x$ is the
coordinate perpendicular to the layers, and $\alpha$ specifies the
orientation of the spin relative to the direction of the
magnetization of the ferromagnet.  We assume that the electrons in
the F layer are described  by a single-band Hubbard model$^{15}$.
In the Hartree-Fock approximation, this model yields one-electron
energies $\varepsilon_{k,\alpha }=\varepsilon_k+I<n_{-\alpha} >$
where $I$ is the on-site screened Coulomb interaction. To simplify
the calculations, we shift the energies in the F-region by
$-I<n\uparrow>$. This brings the zero of energy to the bottom of
the minority band. In this case, the spin-up conduction electron
sees an attractive square well potential of depth $\Delta =
I(<n\uparrow>-<n\downarrow>)$, a quantity defining the exchange
splitting between the majority and minority spin bands. The down-
spin electron moves in a potential that is constant and equal to
zero for all $x$.

Thus, $U_{\downarrow}(x) = 0$ for
 $- \infty< x < \infty$, $U_{\uparrow} (x) = - \Delta$ for $-D/2<
x < D/2, U_{\uparrow}(x) = 0 $ for $x < - D/2$ and $x
>D/2$.

If the zero of energy is taken at the bottom of the majority band,
the up-spin electron sees a potential barrier of height $\Delta$,
while the down-spin electron moves in constant zero-value
potential. This choice of potential has been previously used by
Bruno$^{10,16}$. Mills$^{12}$ assumes that the potential energy of
the up-spin electron is $V_0$ above the zero of energy (set at the
bottom of the conduction band) whereas the down-spin has potential
energy $V_0+\Delta $. Thus, Bruno's potential follows by setting
$V_0=0$ and our choice corresponds to taking $V_0=-\Delta $. The
fact that the results$^{12}$ for the $D$-dependence of $\alpha'$
are not very sensitive to the choice of $V_0$ provides some
justification for the present choice of the potential. Our main
motivation for this choice is that the $x$-component of the
electron wave vector for spin up stays real (see Eq.(9)). This
simplifies the summation over the transverse modes in Eqs. (4) and
(5).

The single-band Hubbard model implies that, owing to the high
density of states, the effective mass of the itinerant electrons
in the ferromagnetic layer is larger than the free electron mass.
This complicates the scattering problem since, besides the
potential well, there is also a "mass barrier" in the F-region.
Similar to Refs. 10 and 16, we replace this position dependent
mass by a constant average mass. It should be pointed out that an
enhancement of the effective mass can also take place in normal
layers due to electron-electron interactions. This effect is
especially pronounced for Pd and Pt layers$^{13}$.

We first consider the reflection amplitudes $r^{(\alpha)}_{m,n}$.
Since the transverse momentum is conserved, we have
$r^{(\alpha)}_{m,n} = r^{(\alpha)}_{m} \delta_{m,n}$ where
$r^{(\alpha)}_{m}= r^{(\alpha)}_{0}(k_{\perp}^{(\alpha)})$ where
$k ^{(\alpha)}_{\perp}$ is the component of the wave vector along
axis $x$. For our choice of the potential, we have
$r^{\downarrow}_{0} (k^{\downarrow}_{\perp}) = 0$, and$^{16}$
\begin{equation}\label{Eq6}
r^{\uparrow}_{0}(k^{\uparrow}_{\perp})=
r^{\uparrow}_{\infty}(k^{\uparrow}_{\perp})\frac{1 - \exp (2 i
k^{\uparrow}_{\perp}D)}{1 -
[r^{\uparrow}_{\infty}(k^{\uparrow}_{\perp})]^{2} \exp (2 i
k^{\uparrow}_{\perp}D)}
\end{equation}
 where
 \begin{equation}\label{Eq7}
r^{\uparrow}_{\infty}(k^{\uparrow}_{\perp})= \frac{k_{\perp}-
k^{\uparrow}_{\perp} }{k_{\perp}+ k^{\uparrow}_{\perp}}
\end{equation}

where $k_{\perp}$ and $k^{\uparrow}_{\perp}$ are the wave vectors
in the N and F-regions, respectively.  Since the expressions (4)
and (5) are to be evaluated at the Fermi energy, we have
\begin{equation}\label{Eq8}
k_{\perp}= (k^{2}_{F} - k^{2}_{||})^\frac{1}{2}
\end{equation}
and
\begin{equation}\label{Eq9}
k^{\uparrow}_{\perp}= (k^{2}_{F} - k^{2}_{||} +
\delta^{2})^\frac{1}{2}
\end{equation}
where $k_{||}$ is the magnitude of the in-plane wave vector
$\vect{k}_{||}$, and $\delta^2 =2m\Delta/\hbar^2$. Eq. (9) implies
that $k^{\uparrow}_{\perp}$ stays real over the whole range, $(0,
k_{F})$, of the variable $k_{||}$.

Next let us consider the transmission amplitudes.  The reflection
symmetry of the potential implies $t
^{'(\alpha)}_{m,n}=t^{(\alpha)}_{m,n}$.  Due to the conservation
of transverse momentum, we have
$t^{(\alpha)}_{m,n}=t^{(\alpha)}_{0}(k^{(\alpha)}_{\perp})\delta_{m,n}$
where $t^{\downarrow}_{0}(k^{\downarrow}_{\perp})= 1$ and
\begin{equation}\label{Eq10}
t^{\uparrow}_{0}(k^{\uparrow}_{\perp})= \frac{1 -
[r^{\uparrow}_{\infty}(k^{\uparrow}_{\perp})]^{2}}{1 -
[r^{\uparrow}_{\infty}(k^{\uparrow}_{\perp})]^{2} \exp (2i
k^{\uparrow}_{\perp} D)}
\end{equation}

\section{ Parameters $ A_r,A_i$}

We are now ready to evaluate the quantities $A_{r}$ and $A_{i}$.
The transverse-mode sums in Eqs. (4) and (5) can be written as a
sum over the in-plane wave vectors $\vect{k}_{||}$.  Converting
the sum to a two-dimensional integral and using Eq. (9), we have
\begin{equation}\label{Eq11}
\sum_{m,n}= \frac{L^{2}}{2 \pi} \int^{k_{F}}_{0}d k_{||} k_{||}= -
\frac{L^{2}}{2 \pi}\int^{k_{2}}_{k_{1}}d k
^{\uparrow}_{\perp}k^{\uparrow}_{\perp}
\end{equation}
where $L$ is the lateral dimension of the film.  Owing to the
presence of the term $\exp (2 i k^{\uparrow}_{\perp} D)$ in Eqs.
(6) and (10), it is convenient to integrate over the variable
$k^{\uparrow}_{\perp}$. According to Eq. (9), the integration
limits for this variable are
\begin{equation}\label{Eq12}
k_{1}= (k_{F}^{2}+ \delta^{2})^\frac{1}{2}
\end{equation}
\begin{equation}\label{Eq13}
k_{2}= \delta
\end{equation}

Using the unitarity property of the scattering matrix, we have
\begin{equation}\label{Eq14}
(r^{\uparrow}_{m,n})^{2}+ (t^{\uparrow}_{m,n})^{2}= 1
\end{equation}
Using this relation and the fact that $r^{\downarrow}_{m,n}=0, t
^{\downarrow}_{m,n}=1$, equation (4) can be simplified to the
following form
\begin{equation}\label{Eq15}
A_{r}= \sum_{m,n} (1 - \textrm{Re} t^{\uparrow}_{m,n})
\end{equation}
Performing the summation with use of the prescription (11) and
using Eq. (10), we obtain from Eq. (15)
\begin{eqnarray}\label{Eq16}
A_{r}= \frac{L^{2}}{2 \pi} \int^{k_{1}}_{k_{2}} d k
^{\uparrow}_{\perp} k^{\uparrow}_{\perp}\nonumber \\* \times
\frac{[(r^{\uparrow}_{\infty})^{2}+
(r^{\uparrow}_{\infty})^{4}][1-\cos(2 k^{\uparrow}_{\perp} D )]}{1
- 2 (r^{\uparrow}_{\infty})^{2}\cos (2 k^{\uparrow}_{\perp} D) +
(r ^{\uparrow}_{\infty})^{4}}
\end{eqnarray}
where

\begin{equation}\label{Eq17}
r^{\uparrow}_{\infty} = \frac{[(k^{\uparrow}_{\perp})^{2} -
\delta^{2}]^{\frac{1}{2}} -
k^{\uparrow}_{\perp}}{[(k^{\uparrow}_{\perp})^{2} -
\delta^{2}]^{\frac{1}{2}} + k^{\uparrow}_{\perp}}
\end{equation}

Substituting $(r^{\downarrow}_{m,n} = 0, t^{\downarrow} _{m,n} =
1)$ into Eq. (5), we obtain
\begin{equation}\label{Eq18}
A_{i} =  \sum_{m,n} \textrm{Im} t^{\uparrow}_{m,n}
\end{equation}

This result can be expressed using Eqs. (10) and (11) as an
integral over $k^{\uparrow}_{\perp}$
\begin{eqnarray}\label{Eq19}
A_{i} = \frac{L^{2}}{2 \pi} \int^{k_{1}}_{k_{2}}
dk^{\uparrow}_{\perp}k^{\uparrow}_{\perp} \nonumber \\* \times
\frac{[(r^{\uparrow}_{\infty})^{2}- (r^{\uparrow}_{\infty})^{4}]
\sin (2 k^{\uparrow}_{\perp} D)}{1 - 2
(r^{\uparrow}_{\infty})^{2}\cos (2 k^{\uparrow}_{\perp} D) + (r
^{\uparrow}_{\infty})^{4}}
\end{eqnarray}

 Making a
 substitution $u = 2D k^{\uparrow}_{\perp}$, we  obtain from
 Eqs. (16) and (17)
 \begin{equation}\label{Eq20}
 A_{r}= \frac{L^{2}\delta^{2}}{2 \pi} I _{r}(u_{1},
 u_{2})
\end{equation}

where
\begin{eqnarray}\label{Eq21}
I_{r} (u_{1}, u_{2}) = \int ^{u_{1}}_{u_{2}} duu \nonumber \\*
\times \frac{(1 - \cos u )\big
\{u_2^4+[(u^{2}-u^{2}_{2})^{\frac{1}{2}}- u]^{4} \big\}} {2 u
^{6}_{2}(1 - \cos u) + 16 u ^{2}_{2}u^{2}(u^{2} - u ^{2}_{2})}
\end{eqnarray}

According to Eqs. (12) and (13), the dimensionless parameters
$u_{1}$ and $u_{2}$ are given by
\begin{equation}\label{Eq22}
u_{1} = 2 D k _{1}= 2 D (k ^{2}_{F}+ \delta ^{2})^{\frac{1}{2}}
\end{equation}
\begin{equation}\label{Eq23}
u_{2}= 2D k_{2} = 2D \delta
\end{equation}

With the same substitution, Eq. (19) yields
\begin{equation}\label{Eq24}
A_{i} =- \frac{L^{2}\delta ^{2}}{2 \pi}I_{i}(u_{1}, u_{2})
\end{equation}
where
\begin{eqnarray}\label{Eq25}
I_{i} (u_{1}, u_{2}) = \int ^{u_{1}}_{u_{2}} duu \nonumber \\*
\times \frac {\sin u \big \{[(u^{2}-u_2^2)^{\frac{1}{2}}-
u]^{4}-u_2^4 \big\}} {2 u ^{6}_{2}(1 - \cos u) + 16 u
^{2}_{2}u^{2}(u^{2} - u ^{2}_{2})}
\end{eqnarray}

Integrals (21) and (25) are evaluated numerically as a function of
the parameter $u_{2}= 2 D \delta$ by setting $u_{1}/u_{2}=2.38$.
This value results from Eqs. (22) and (23) by taking
$\varepsilon_F=7\textrm{eV}$ and $\Delta=1.5\textrm{eV}$ as
adopted by Bruno$^{10}$ for the Co/Cu/Co system. We note that this
set of parameters corresponds to replacing the $x$-dependent
electron mass by the free-electron mass.

 The results are plotted in Figs. 1
and 2. The data are given for the range $0<2D\delta< 60$. Taking
$k_{F}\simeq\pi/a$ where $a$ is the lattice constant of the
ferromagnetic film,  the upper limit of this range corresponds to
$D_{max}\simeq 20a$. From Fig. 1 we see that, for large values of
$D$, the integral $I_{r}(D)$ tends to a constant equal to $0.08$.
As $I_r(D)$ approaches this value, it exhibits large quantum
oscillations with an initial amplitude of $0.12$. The period of
these oscillations is about $\pi/\delta$. This result can be
traced to the singular behavior of the integrands of Eqs. (21) and
(25). In fact, a numerical plot of the coefficient of the factor
$1-\cos u$ in the integrand of Eq. (21) shows a sharp peak at
$u_2\simeq 2D\delta$. Also the coefficient of the factor $\sin u$
in the integrand of Eq. (25) shows a peak at this value of $u_2$.
This behavior of the integrands is also responsible for the fact
that the integrals (21) and (25) are insensitive to changes of the
ratio $u_1/u_2$. For instance, on increasing this ratio from 2.38
to 10, the plots of the $I_r$ and $I_i$ functions shown in Figs. 1
and 2 sustain less than a 10$\%$ change.

 The amplitude envelope of the oscillations of the function $I_r$
 decreases slowly with increasing
$D$ as a power law $D^{-n}$ where the exponent $n \simeq 0. 6 \pm
0.1$.

The integral (25), displayed in Fig. 2 as a function of the
parameter $2d \delta $, exhibits oscillations of similar character
taking place about $I_i(D)=0$. The position and magnitude of the
first peak of the function $I_i(D)$ roughly agree with those of
the oscillatory part of the $I_r(D)$ curve.

\section{Gilbert damping}

The excess damping constant $\alpha'$ is given by the second term
on the right hands side of Eq. (3). Assuming that the increment
the gyromagnetic ratio is small, we set $\gamma/\gamma_0\simeq 1$
and obtain with use of Eq. (20)

\begin{equation}\label{Eq26}
\alpha'\simeq \frac{g_L \mu_B A_r}{2\pi M_s L^2
D}=\frac{g_L\mu_B\delta^2 I_r(D)}{4\pi^2M_s D}
\end{equation}

 where $M_s$ is the saturation magnetization of the ferromagnetic
layer.

 Invoking Fig. 1, we see from Eq. (26) that, for $D\gg
\delta^{-1}$, $\alpha'(D)$ involves a $D^{-1}$ dependence similar
to previous works$^{1,4,13}$. Superimposed on this dependence are
large slowly decaying short-wave oscillations with a period close
to $\pi/\delta$. In the limit of $D\gg 1/\delta$, the N/F and F/N
interfaces are decoupled, and presumably
 contribute to $\alpha'$ independently$^{4,13}$. In Ref. 13, the
interfaces were represented by monolayers acting as delta function
scatterers. The expression for $\alpha'$ obtained in this
approximation is, to leading order in $J_{sd}$, proportional to
$J_{sd}^2 \propto \Delta^2$. This contrasts with Eq. (26) showing
a linear dependence on $\Delta$. A comparison of this result with
the dynamic RKKY theory of Mills$^{12}$ is made in Sec. VI.

Let us now estimate the magnitude of $\alpha'$ for the Cu/Co/Cu
system. Consistent with the choice $u_1/u_2=2.38$ made in the
evaluation of the integrals (21) and (25), we take $\delta^2\simeq
2m\Delta/\hbar^2$ where $\Delta$=1.5 eV. Letting $M_s\simeq 1400$
G, $D\simeq 40\AA$ and $I_r(D)\approx 0.1$, we obtain from Eq.
(26) $\alpha'\simeq 3\times 10^{-4}$. This is possibly an
underestimate since a free-electron mass is assumed$^{10}$. An
enhancement of the average mass due to high density of states in
the Co layer can increase the value of $\delta^2$ in Eq. (26)
yielding $\alpha'$ of order $10^{-3}$.

Heinrich et al.$^{17}$ studied, using FMR, epitaxially stabilized
ultrathin $\textit{fcc}$ films of Co deposited on bulk Cu(001).
For a 10 monolayer (ML) Co film, they find $\alpha \approx
10^{-2}$. Schreiber et al.$^{18}$ measured Gilbert damping of
single crystal $\textrm{Fe}_x\textrm{Co}_{1-x}$ alloy films
prepared by sputtering on MgO(001) substrates. For pure Co films
$200\AA$ thick, they measure $\alpha\approx 10^{-2}$ for the hard
and $\alpha\approx 6\times 10^{-3}$ for the easy direction,
respectively. On the other hand, the pure Fe films show
substantially lower damping $(\alpha\approx 2\times 10^{-3})$ that
is practically isotropic$^{18}$. On comparing our predicted
damping with the experimental results on Co films$^{17,18}$, it is
evident that the magnitude of the oscillatory part of $\alpha$ is
within the error margins of the measured damping.

Another way to estimate the Gilbert damping constant is to explore
the current-induced switching of magnetic moments in Co/Cu/Co
systems. Katine et al.$^{19}$ used thin-film pillars containing a
thin (25$\AA$) Co film coupled to a thicker (100$\AA$) Co film.
During the rotation of the magnetization of the thinner film, the
magnetization of the thicker film remains fixed. This enables the
determination of the polarity of the current bias associated with
the spin-transfer excitations in the thinner layer. Moreover, the
thicker layer acts as a spin sink. In this way, problems with spin
accumulation in the Cu layers can be circumvented$^4$. The Gilbert
damping constant is determined starting from the Landau-Lifshitz
equation augmented by the current-induced spin transfer
torque$^{20}$. The stability of the given magnetic configuration
of the pillar is determined by a competition between the
spin-transfer torque and the Gilbert damping torque. From the way
the critical switching current scales with the effective magnetic
field, Katine et al.$^{19}$ find that an agreement with their
experiment is obtained by assuming $\alpha\approx 7\times
10^{-3}$. This value is in good agreement with the FMR
data$^{17,18}$. The current-induced switching has been also
studied by injecting the current into the multilayer through a
point contact$^{21}$. Interestingly, the damping constant deduced
by this method is 10-50 times larger than the value obtained in
the pillar geometry$^{19}$. This could be related to the fact
that, in the point-contact study, the spin-wave excitation is
induced in a localized region exchange coupled to the unbounded
ferromagnetic film.

  Due to the relatively small FMR linewidth of Fe films$^{18}$,
  multilayers involving Fe layers seem more suitable for the
  observation of the quantum oscillations. Urban et al.$^2$ studied
  FMR on two Fe layers separated by nonmagnetic Au spacer. The thickness  of
  the thinner layer is varied from 8 to 31ML. The
  thicker, 40ML, layer exhibits precession of
  negligible amplitude and serves as a spin sink for the spin
  currents emitted form the precessing thin layer. The FMR linewidth
  follows
  a $D^{-1}$ dependence without superimposed oscillatory component. For the 16ML film,
  the measured excess Gilbert damping constant, $\alpha'\approx 2\times 10^{-3}$, is comparable to
  the intrinsic damping in the single Fe film$^{18}$.

  Using Eq. (26), we estimate the theoretical value
  $\alpha'_{th}$ for the 16ML Fe film imbedded between Au layers. We assume that
  the electronic parameters are similar to those of the Fe/Cu/ Fe
  system discussed by Hood and Falicov$^{22}$. The
  conduction electron mass is taken, by these authors, independent of the material and the
  spin orientation and equal to $ m^{*}\approx 4\times \textrm{free electron
  mass}$. The exchange splitting for Fe is$^{22}$  $\Delta\approx 2.5$ eV. The resulting value of the parameter
  $\delta^2=2m^{*}\Delta/\hbar^2$ is $2.6\times
  10^{16}\textrm{cm}^{-2}$. Consequently, the period of the
  oscillations of the functions $I_r$ and $I_i$ is
  $\pi/\delta \approx 2\times 10^{-8}\textrm{cm}$. Since this period
  is comparable to the lattice constant of Fe,  films differing in
  thickness by 1ML should be investigated to detect the
  oscillations. Using this value of $\delta^2$, and taking
  $M_s\approx 1700 G$, $D\approx 46
  \AA$ and $I_r\approx 0.1$, we obtain from Eq. (26) the excess damping
  constant $\alpha'_{th}\approx 1.7\times 10^{-3}$. This is
  comparable with the experimental value$^2$.

  \section{Gyromagnetic ratio}

Next we consider the renormalization of the gyromagnetic ratio
$\gamma$. Putting $\gamma =\gamma_0 +\Delta\gamma$ where
$\Delta\gamma\ll \gamma_0$, we obtain from Eq. (2)

\begin {equation}\label{Eq27}
\frac{\Delta\gamma}{\gamma_0}=\frac{\Delta
g}{g_0}\simeq-\frac{g_L\mu_B}{2\pi M}A_i
\end{equation}

where $\Delta  g=g-g_0$ is the change of the $g$-factor due to the
spin-pumping mechanism.

 Using Eqs. (20), (24), (26) and (27), we evaluate the ratio

\begin{equation}\label{Eq28}
\frac{\Delta g}{g_0\alpha'}=
-\frac{A_i}{A_r}=\frac{I_i(u_1,u_0)}{I_r(u_1,u_2)}
\end{equation}

From Figs. 1 and 2 we see that the oscillatory part of $I_r$ is of
similar form and magnitude as $I_i$. This implies, that an
observation of quantum oscillations of $\alpha'$ should be
accompanied by an oscillation, of similar magnitude and period, of
the ratio $\Delta g/g_0$.

  Fig. 2 implies that for thicker films, such that
quantum coherence is suppressed, the shift of the $g$-factor
should be negligible compared to the measured value of $\alpha'$.
This prediction is in disagreement with the FMR data on Pd/Py/Pd
and Pt/Py/Pt where Py is a permalloy film$^1$. These data yield a
damping parameter $\alpha'$ that is proportional to $1/D$ and does
not exhibit any superimposed quantum oscillations. This suggests
that the trilayers are in a quantum-incoherent regime for which
the present theory predicts $\Delta g/g_0 \simeq 0$. In contrast,
the experimental value of this quantity is comparable with the
experimental damping constant $\alpha'$. For instance, for the
Pd/Py/Pd system with $D=50\AA$, $\alpha'\approx \Delta
g/g_0\approx 10^{-2}$. Thus we have
\begin{equation}\label{Eq29}
\biggr [\frac{\Delta g}{g_0\alpha'}\biggr ]_{exp}\approx 1
\end{equation}

It is interesting that an order of magnitude agreement with this
result can be obtained by evaluating the ratio $\Delta
g/(g_0\alpha')$ from the model of two monolayers represented by a
delta function potential$^{13,14}$. For Pd and Pt, the Coulomb
interactions between the electrons need to be taken into account.
In Ref. 14 we have shown that the effect of these interactions is
to enhance the delta function potential of the monolayer by the
Stoner factor $S_E$. As a result, the quantity $A_i$ for the
monolayer is changed to$^{14}$

\begin{eqnarray}\label{Eq30}
\widetilde{A_i}\approx \frac{1}{\pi}L^2k_F\beta S_E \nonumber \\*
\times \biggr [1+\frac{\beta^2 S_E^2}{k_F^2+\beta^2
S_E^2}-\frac{2\beta S_E}{k_F}\arctan\frac{k_F}{\beta S_E}\biggr ]
\end{eqnarray}

where $\beta\approx k_F J_{sd}/(2\varepsilon_F)$. In what follows,
we assume that $J_{sd}$ is negative$^{23}$. The origin of this
antiferromagnetic s-d interaction is quantum-mechanical mixing
between conduction and localized $d$-orbitals. In Sec.II we found
that the single-band Hubbard model$^{15}$ yields a ferromagnetic
interaction between the itinerant electron and the magnetization.
This results from the Hartree-fock approximation which linearizes
the on-site Coulomb interaction of two electrons with opposite
spin. Thus the energy of the up-spin electron is lower relative to
the down-spin one. There are no consequences of this dichotomy for
the quantity $\alpha'$. However, the \textit{sign} of $\Delta g$
changes as the interaction goes from the antiferromagnetic to the
ferromagnetic one.

The quantity $A_r$ for the monolayer becomes$^{14}$

\begin{equation}\label{Eq31}
\widetilde{A_r}\approx\frac{1}{\pi}L^2 \beta^2 S_E^2 \biggr [
\ln(\frac{k_F^2}{S_E^2\beta^2}+1) -\frac{k_F^2}{k_F^2+\beta^2
S_E^2}\biggr ]
\end{equation}
Setting $J_{sd}\approx -\varepsilon_F/10$ and $S_E\approx 5$, Eqs.
(30) and (31) yield

\begin{equation}\label{Eq32}
\frac{\Delta g}{g_0\alpha'}\approx
-\frac{\widetilde{A_i}}{\widetilde{A_r}}\approx 0.8
\end{equation}

Applying this single-monolayer result to the model of two
monolayers located at the N/F and F/N interfaces$^{13}$, the same
result is obtained for the ratio $\Delta g/(g_0\alpha')$. We see
that the theoretical result (32) is in order of magnitude
agreement with Eq. (29).

\section{Comparison to dynamic RKKY theory}

We now compare the results of Sec. IV and V to the dynamic RKKY
theory of Mills$^{12}$. Let us begin by considering the excess
Gilbert damping given by Eq. (26). We note that the formalism of
the spin-pumping approach is very different from the dynamic RKKY
theory$^{13,14}$. Thus we limit ourselves to comparing only the
final results.

 Using the relation $G'=\gamma M_s \alpha'$, we obtain from
 Eq.(26) the excess Gilbert damping constant

 \begin{equation}\label{Eq33}
 G'\simeq\frac{\mu_B^2\delta^2 I_r(D)}{\pi^2 \hbar D}
 \end{equation}

 We wish to compare this equation with the excess damping constant obtained using the dynamic
  RKKY method (see  Eq.(29) of Ref. 12)

 \begin{equation}\label{Eq34}
 \Delta G=\frac{\mu_B^2
 k^{\uparrow}(\varepsilon_F)k^{\downarrow}(\varepsilon_F)f(D)}
 {8\pi^2 \hbar D}
 \end{equation}

 where $f(D)$ is a dimensionless function which, for $k^{\downarrow}(\varepsilon_F)D\gg
 1$, approaches a constant not far from unity. This function
 exhibits quantum oscillations that consist of short-wave
 oscillations with period
 $2/[k^{\uparrow}(\varepsilon_F)+k^{\downarrow}(\varepsilon_F)]$
 superposed on long-wave oscillations with period
 $2/[k^{\uparrow}(\varepsilon_F)-k^{\downarrow}(\varepsilon_F)]$.
 In contrast, the function $I_r(D)$ exhibits only short-wave oscillations
 with period $\pi/\delta$. It approaches, as $D$ increases,
 a constant close to 0.1.

  As far as the magnitude of the damping
 is concerned, the main difference between Eqs. (33) and (34) is
 that the factor
 $k^{\uparrow}(\varepsilon_F)k^{\downarrow}(\varepsilon_F)$ is
 replaced by $\delta^2$. This implies that the magnitude of
 $G'$, for large $D$, is reduced compared to $\Delta G$ by a
 factor of order $\Delta/\varepsilon_F$. A more serious
 discrepancy is that Eq. (34) yields a nonzero damping even
  when the splitting of the up and down-spin bands vanishes
  (
 $k^{\uparrow}(\varepsilon _F)=k^{\downarrow}(\varepsilon_F)$).
 According to Eq. (4), $A_r$ vanishes in this case since the
 reflection and transmission amplitudes for the up-spin cancel
  those for the down-spin electron. This feature of the
  spin-pumping theory has been emphasized by Tserkovnyak \textit{et al.}$^4$ when comparing
  their formula for excess damping with the result of Berger$^3$.
  The requirement of nonzero spin-splitting also
   follows from the general dynamic RKKY approach$^{13}$ to
   Gilbert damping. We return to a possible resolution of this
   difficulty at the end of this section.

 In what follows, we argue that the absence of long-wave
 oscillations in $I_r(D)$ is an upshot of the simplified
 potential ($U_{\downarrow}(x)=0$). Similar argument has been raised
 by Bruno$^{10}$ for the interlayer exchange coupling. When both
 the up and down-spin electrons are scattered, a long-wave modulation of the
 short-wave oscillation is expected for the $D$-dependence of this
 coupling.

 To substantiate this idea, we derive
 $A_r$ for the case when both $U_{\downarrow}(x)$ and
 $U_{\uparrow}(x)$ are nonzero. Using Eqs. (6-10), we obtain from
 Eq. (4)

\begin{eqnarray}\label{Eq35}
A_r\simeq \frac{L^2}{2\pi} \int^{k_F}_{0}dk_{\|}k_{\|}
(B^{\uparrow} B^{\downarrow})^{-1} \biggr \{B^{\uparrow}
B^{\downarrow}
\\*\nonumber -[1-(r_{\infty}^{\uparrow})^2][1-(r_{\infty}^{\downarrow})^2]
-r_{\infty}^{\uparrow}r_{\infty}^{\downarrow} \\*\nonumber \times
[1+(r_{\infty}^{\uparrow})^2
+(r_{\infty}^{\downarrow})^2+(r_{\infty}^{\uparrow}r_{\infty}^{\downarrow})^2]\\*\nonumber
+\{(r_{\infty}^{\uparrow})^2[1-(r_{\infty}^{\uparrow})^2][1-(r_{\infty}^{\downarrow})^2]
\\*\nonumber +r_{\infty}^{\uparrow}r_{\infty}^{\downarrow}[1+2(r_{\infty}^{\uparrow})^2+
(r_{\infty}^{\uparrow}r_{\infty}^{\downarrow})^2]\}\\*\nonumber\times
\cos(2k_{\perp}^{\uparrow}D)+\{r_{\infty}^{\uparrow}\rightleftarrows
r_{\infty}^{\downarrow}\}\cos(2k_\perp ^{\downarrow}D)
\\*\nonumber-\{r_{\infty}^{\uparrow}r_{\infty}^{\downarrow}[(r_{\infty}^{\uparrow})^2
+(r_{\infty}^{\downarrow})^2]\}\cos[2(k_{\perp}^{\uparrow}+k_{\perp}^{\downarrow})D]
\\*\nonumber-\{r_{\infty}^{\uparrow}r_{\infty}^{\downarrow}[1+(r_{\infty}^{\uparrow}
r_{\infty}^{\downarrow})^2]+(r_{\infty}^{\uparrow}r_{\infty}^{\downarrow})^2
[1-(r_{\infty}^{\uparrow})^2]\\*\nonumber\times
[1-(r_{\infty}^{\downarrow})^2]\}\cos[2(k_{\perp}^{\uparrow}-k_{\perp}^{\downarrow})D]\biggr
\}
\end{eqnarray}

where
\begin{equation}\label{Eq36}
B^{(\alpha)}=1-2(r_{\infty}^{(\alpha)})^2
\cos(2k_{\perp}^{(\alpha)}D)+(r_{\infty}^{(\alpha)})^4
\end{equation}

Generalizing Eq. (9) to both spin components, we set

\begin{equation}\label{Eq37}
k_{\perp}^{(\alpha)}=[(k_F^{(\alpha)})^2-k_{\parallel}^2]^{\frac{1}{2}}
\end{equation}

where $k_F^{(\alpha)}=(k_F^2+\delta_{\alpha}^2)^{\frac{1}{2}}$ is
to be identified with $k^{(\alpha)}(\varepsilon _F)$ of Ref. 12.

 To determine the periods of the oscillations of $A_r(D)$, we consider
the asymptotic estimate of the integral (35) for $D$ large. We
note that the major contribution to this integral arises from the
immediate vicinity of the end points of the integration interval
$(0,k_F)$ and from the vicinity of the stationary points of the
function $k^{(\alpha)}_{\perp}$. From Eq. (37) we see that the
derivative of this function vanishes at the stationary point
$k_{\parallel}=0$. However, the contribution of this stationary
point to the integral (35) vanishes owing to the factor
$k_{\parallel}$ which causes the coefficients of all cosine terms
to vanish.

 For the same reason, the partial integration method
yields a nonzero contribution only from the end point
$k_{\parallel}=k_F$. This conclusion holds also for the simplified
potential used in Sec. III. In fact, setting
$r_{\infty}^{\downarrow}=0$, Eq.(35) reduces to the integral (16)
which is dominated by vicinity of
$k_{\perp}^{\uparrow}(k_{\parallel}=k_F)=\delta $.

Consequently, Eq. (35) yields three short-wave oscillatory terms
with periods $\pi/\delta_{\uparrow}, \pi/\delta_{\downarrow},
\pi/(\delta_{\uparrow}+\delta_{\downarrow})$ and a long-wave term
with period $\pi/(\delta_{\uparrow}-\delta_{\downarrow})$. Hence,
once electrons of both spin orientations are allowed to scatter,
the spin-pumping approach exhibits oscillations similar to those
of Ref. 12. The main distinction is that the wave vectors
$k^{(\alpha)}(\varepsilon_F)$ determining the periods of the
oscillations in the dynamic RKKY theory$^{12}$ are replaced by the
wave vectors $ \delta_{\alpha}$.

We now return to the aforementioned problem concerning the product
$k^{\uparrow}(\varepsilon_F)k^{\downarrow}(\varepsilon_F)$ in Eq.
(34). First, we examine the integral (35) for the case when
$\delta_{\uparrow}=\delta_{\downarrow}$ (implying
$r_{\infty}^{\uparrow}=r_{\infty}^{\downarrow}$). In this case, it
is possible to verify explicitly that $A_r(D) \rightarrow 0$ as
$D\rightarrow\infty $. Specifically, the rapid oscillations tend
to cancel the contribution of the three short-wave cosine terms in
the integrand. The remaining terms, upon substituting
$r_{\infty}^{\uparrow}=r_{\infty}^{\downarrow}$, exactly cancel so
that the integral (35) vanishes. This is in accord with the more
general formula (4). Moreover, it provides an independent check
that the rather complicated integrand of Eq. (35) is correct.
Based on this result and the fact that, upon letting
$r_{\infty}^{\downarrow}=0$, this equation goes over to Eq. (16),
we propose the following generalization of Eq. (20)

\begin{equation}\label{Eq38}
A_r(D)\propto L^2 (\delta_{\uparrow}-\delta_{\downarrow})^2 F_r(D)
\end{equation}

where $F_r(D)$ tends to a constant for $D$ large and contains a
superposition of slowly decaying oscillations with periods
$\pi/\delta_{\uparrow},\pi/\delta_{\downarrow},\pi/(\delta_{\uparrow}+\delta_{\downarrow})$
and $\pi/(\delta_{\uparrow}-\delta_{\downarrow})$. Eq. (38) also
suggests that replacing the product
$k^{\uparrow}(\varepsilon_F)k^{\downarrow}(\varepsilon_F)$ by
$[k_{\uparrow}(\varepsilon_F)-k_{\downarrow}(\varepsilon_F)]^2$
may endow Eq. (34) with a correct dependence on the splitting of
the up and down-spin electron bands.

In the preceding section we came to the conclusion that quantum
oscillations of the damping constant $\alpha'$ are accompanied by
oscillations of similar magnitude and period of the relative
shift, $\Delta g/g_0$, of the gyromagnetic ratio $g$. This
spin-pumping result disagrees with that of the dynamic RKKY
method$^{12}$. In this method, the renormalization of $g$ stems
from the frequency derivative, called $\Lambda_1$, of the
transverse dynamical susceptibility. Though Ref. 12 points out the
general importance of this quantity in the spin dynamics, it
reports rough estimates indicating that the renormalization
effects due to $\Lambda _1$ are small.

\section{Discussion}

In the itinerant model of a bulk ferromagnet, the $g$-shift and
the damping $\alpha'$ have been attributed to the combined effect
of spin-orbit interaction and the electron-lattice
interaction$^{24}$. Mizukami et al.$^1$ assume that $\Delta g$ and
$\alpha'$ observed on Pd/Py/Pd and Pt/Py/Pt trilayers are due to
similar interactions taking place at the interfaces.

 The results outlined in Eqs. (30)-(32)
represent a strong departure from these ideas. In the framework of
the spin-pumping theory, the enhancement of the gyromagnetic ratio
comes from the angular momentum of the conduction electron spin
carried away from the F-film into the N-layers. It is the spin
current that is in the phase with the precession that contributes
to the $g$-factor. The out of the phase part gives rise to the
excess damping $\alpha'$.

Now in order that this mechanism works, it is necessary that the
conduction electron spins are at equilibrium with the normal metal
reservoirs. Otherwise, spin accumulation takes place which blocks
the flow of the spin current$^4$. The equilibrium is established
by spin-lattice relaxation in the normal metal$^{25}$. This
relaxation process involves the spin-orbit interaction combined
with the electron-lattice interaction similar to the mechanism of
Ref. 24. In the spin-pumping approach, the conduction electron
spin is first transported into the normal metal where it relaxes.
Thus we may view the relaxation of the ferromagnetic
magnetization, in this approach, as a nonlocal process mediated
via spin current. On the other hand, the itinerant electrons of
the bulk ferromagnet relax via local spin-orbit
interaction$^{24}$. Clearly, the spin-orbit interaction is an
inevitable component for magnetization relaxation in both
mechanisms$^{4,24}$.

It remains to address the reasons for the absence of quantum
oscillations in the FMR experiments$^{1,2}$ which study the
Gilbert damping and the $g$-factor as a function of the thickness
of the ferromagnetic film. It is, of course, possible that another
mechanism [see Ref. 1] intervenes and obliterates the predictions
of Eqs. 26 and 27.

 Another possibility is that the oscillations
are washed out by interface roughness. Note that the wavelength of
the short-wave oscillations, considered here, is about 5$\AA$ for
Cu/Co/Cu and only about 2$\AA$ for Cu/Fe/Cu. It is conceivable
that roughness introduces fluctuations of $D$ larger than these
wavelength. Averaging over these fluctuations strongly suppresses
the amplitude of the oscillations. Consequently, one essentially
sees an average which is a constant of order 0.1 for Fig.1, and
zero for Fig. 2. This is in apparent disagreement with the
monolayer model invoked in Sec. V to explain the large shift of
the $g$-factor observed on Pd/Py/Pd systems$^1$. To clarify this
point, we note that the monolayer model assumes that the quantum
oscillations in the ferromagnetic film rapidly decay with the
distance from the interface. In the extreme case of overdamping,
the function $I_i(D)$ does not change sign upon increasing $D$ by
one wavelength form the origin. Thus fluctuations of the thickness
do not average the $g$-shift to zero in the monolayer model.

Though short-wave oscillations are expected to be washed out by
surface roughness, there are also long-wave terms in the
expressions for the Gilbert damping and the shift of the
gyromagnetic ratio that may survive the averaging over $D$. As
seen from Eq. (35), these terms appear in $A_r$ when electrons of
both spin orientations are subject to scattering. It can be shown
in a similar way that these terms are also present in the
expression for $A_i$. Since the period of these oscillations is
$\pi/(\delta_{\uparrow}-\delta_{\downarrow})$, multilayers with
small exchange splitting should be used to increase the chance of
observing the long-wave oscillations.

An alternative way of observing the quantum oscillations is
suggested by the fact that $I_r$ and $I_i$
 are functions of the product $2D\delta$. Thus, instead of changing the
thickness of the ferromagnetic film, we may vary the parameter
$\delta$ while keeping $D$ fixed. For the case when both
$U_{\uparrow}$ and $U_{\downarrow}$ are nonzero, we have
$\delta^2_{\alpha}=2m \Delta_{\alpha} /\hbar^2$ where
$\Delta_{\alpha}$ is the depth of the potential well for electron
of spin $\alpha$. These potential wells can be varied by shifting
the chemical potentials, $\mu_{\alpha}$, of the electrons in the
N-region.

 One possibility is to apply a potential difference
across the N/F interface. This would change the parameters $\delta
_{\uparrow}$ and $\delta_{\downarrow}$ by the same amount.
Consequently, only short-wave oscillations could be tuned in this
manner. It is worth noting that, for the N/F interfaces considered
here, a voltage drop of order 1eV is needed to sweep through
several periods of the short-wave oscillations. To minimize Joule
heating associated with such a voltage drop, the interface
resistance should be large. This could be achieved by a thin layer
of oxide incorporated into the interface.

Tuning of long-wave oscillations requires that the difference
$\Delta_{\uparrow}-\Delta_{\downarrow}$ is varied. It is tempting
to consider the shift
$\Delta\mu=(\mu_{\uparrow}-\mu_{\downarrow})$ due to the spin
polarized current normal to the interface$^{26}$. Using Eq. 12 of
Ref. 26, we estimate that the current density required to generate
a shift of 1eV is of order $10^{10}\textrm{A cm}^{-2}$ which rules
out this idea.

It is of interest, both experimentally and theoretically, to
examine other processes contributing to the decay of multiple
reflections in ideal multilayers. Recent work of Stiles and
Zangwill$^{27}$ is of relevance in this context. These authors
show that the transverse [i.e. perpendicular to $\vect m$]
component of the spin current that flows from the normal metal
into the ferromagnet is absorbed within a distance of several
lattice constants. In the spin-pumping approach$^4$, it is also
the transverse spin current that determines the Gilbert damping
and the gyromagnetic ratio. The parameters $A_r$ and $A_i$ give
the magnitude of these currents. Now, the present calculation of
these parameters shows slowly-decaying oscillations in
disagreement with Ref. 27. Also the numerical results of
Mills$^{12}$ do not show fast decaying oscillations of Gilbert
damping.  We note that Stiles and Zangwill$^{27}$ have identified
three distinct processes contributing to the absorption of the
transverse spin current: (1) spin dependent reflection and
transmission; (2) rotation of reflected and transmitted spins; and
(3) spatial precession of spins in the ferromagnet. Out of these
processes, it is the first one that the present theory and that of
Ref. 12 takes into account. It should be interesting to
incorporate the other two processes into these theories to see if
a fast decay of the quantum oscillations ensues.

 It should be emphasized that these decay processes do not affect
 the longitudinal component of the spin current$^{27}$.  Incidentally, it is also this
 component which determines the exchange coupling between
 ferromagnetic films separated by normal metal spacer. Thus, we
 expect that the $D$-dependence of the the coupling exhibits slowly
 decaying oscillations. This agrees with the theory of
 Bruno$^{10}$ and its experimental verification$^{11}$.

  In view of the predicted fast decay of the transverse spin current, the
$D$-dependent oscillations may be
 decaying faster than calculated in the present simplified model, and
 therefore less easy to detect than the oscillations of exchange
 coupling. In this context, it is worth noting that the decay
 of the transverse spin current depends on the Fermi surface
  mismatch between the ferromagnetic and normal metal Fermi
  surfaces$^{27}$. A slower decay is predicted when the mismatch
  is small $(k^{\uparrow} _F/k_F\approx k^{\downarrow}/k_F)$.
  From this point of view, multilayers with small exchange
  splitting are preferable in order to mitigate the effect of the decay
  processes$^{27}$ on quantum oscillations of the Gilbert damping
  and gyromagnetic ratio.

 \section{Acknowledgement}
 This work was motivated by illuminating discussions with Professor B.
 Heinrich to whom I want to express gratitude. Thanks are also due to Professor D.L.Mills for
 making a copy of Ref. 12 available prior to publication.

\pagebreak
\begin{figure}
\caption{Integral $I_{r} (u_{1}, u_{2})$ defined in Eq. (21). The
plot is obtained by numerical integration as a function of $ u_2=2
\delta D$ for $u_{1}/u_{2}=2.38$. $D$ is the thickness of the
ferromagnetic layer, and $\delta=(2m\Delta/\hbar ^{2})^{1/2}$
where $\Delta$ is the depth of the potential well. }
\label{Fig.1.}
\end{figure}

\begin{figure}
\caption{Plot of the quantity $I_{i}(u_{1},u_{2})$  evaluated from
Eq. (25) as a function of $2 \delta D$ for $u_{1}/u_{2}= 2.38$.}
\label{Fig.2.}
\end{figure}

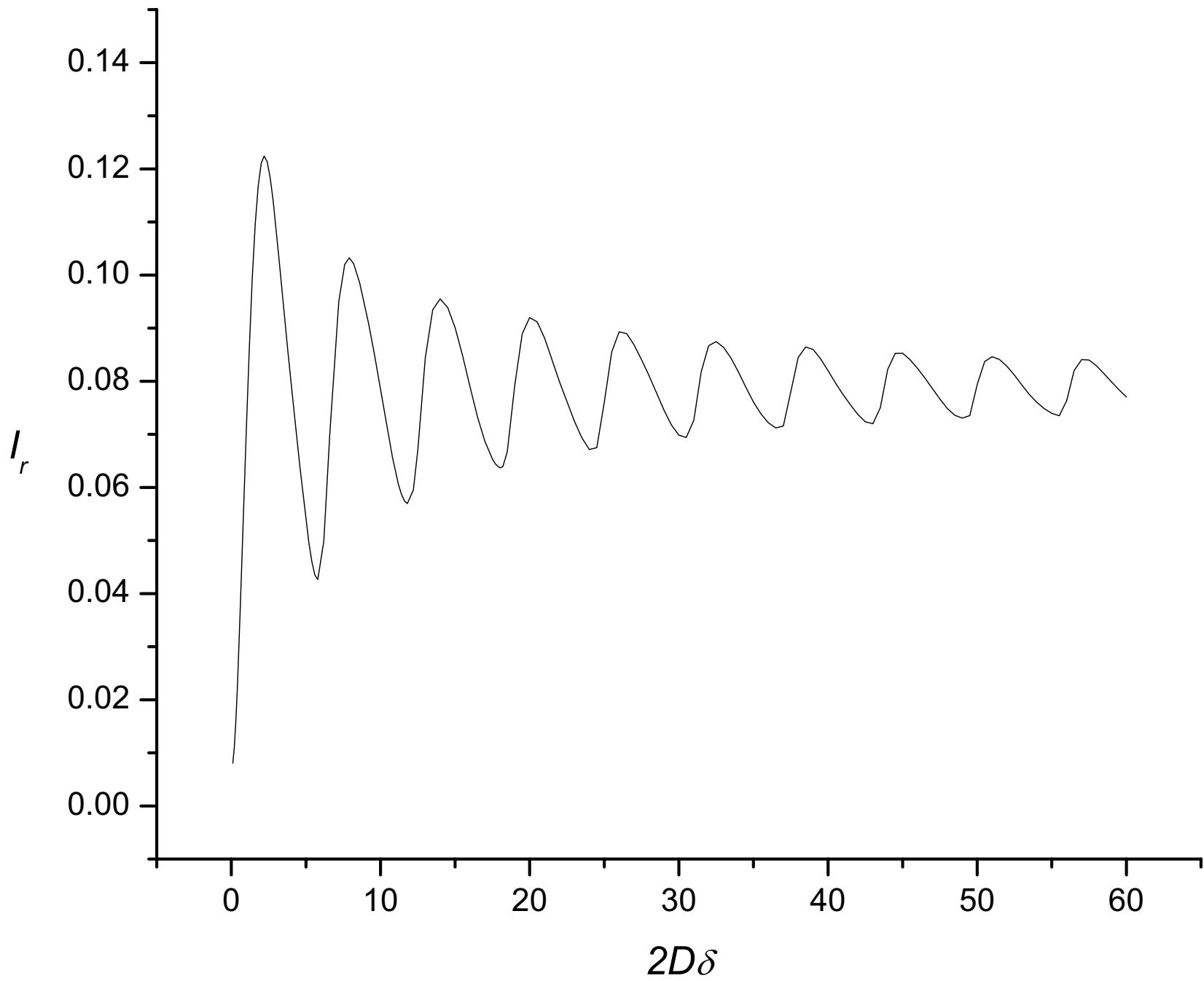
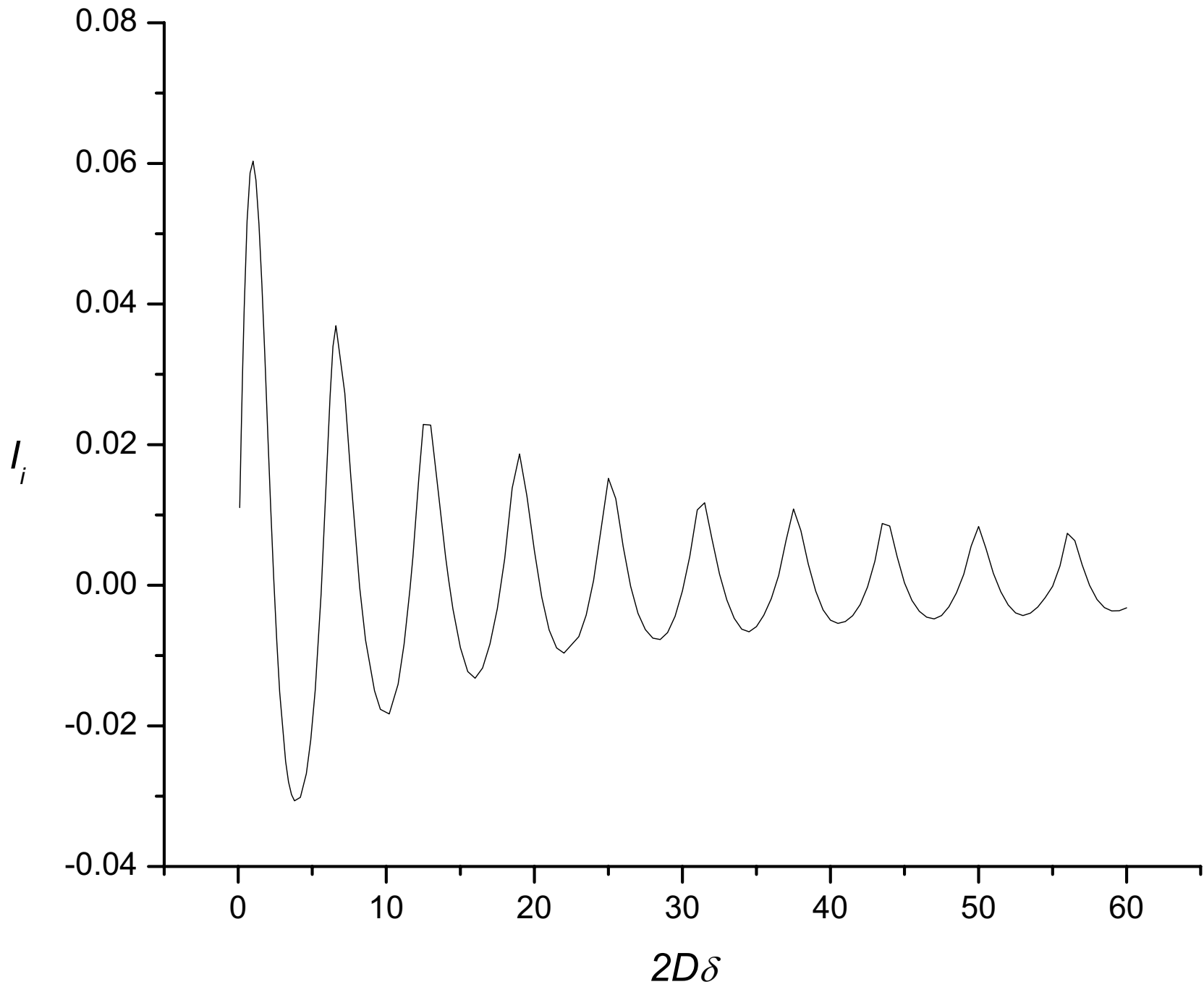
\end{document}